\documentclass{article}

\usepackage{amsmath}

\usepackage[nonatbib,preprint]{neurips_2021}

\usepackage[numbers]{natbib}
\usepackage{nicefrac}       
\usepackage{microtype}      
\usepackage{xcolor}         
\usepackage[pdftex]{graphicx}
\usepackage{xcolor}
\usepackage{bbding}
\usepackage{multirow}
\usepackage{wasysym}
\usepackage{amssymb}
\usepackage{enumitem}
\usepackage{algorithmic}
\usepackage{algorithm}

\usepackage{hyperref}       
\usepackage{color}
\title{Neural dSCA: demixing multimodal interaction among brain areas during naturalistic experiments }

\author{%
  Yu Takagi\thanks{Corresponding author.} \\
  University of Tokyo, Japan \\
  \texttt{yutakagi322@gmail.com} \\
   \And
  Laurence T. Hunt\\
  University of Oxford, UK \\
  \texttt{laurence.hunt@psych.ox.ac.uk} \\
   \And
  Ryu Ohata\\
  University of Tokyo, Japan \\
  \texttt{ryu.oohata@gmail.com} \\
   \And
  Hiroshi Imamizu\\
  University of Tokyo/ATR, Japan\\
  \texttt{imamizu@l.u-tokyo.ac.jp} \\
   \And
  Jun-ichiro Hirayama \\
  AIST, Japan\\
  \texttt{junichiro.hirayama@aist.go.jp} \\
}

\begin{document}

\maketitle

\begin{abstract}
Multi-regional interaction among neuronal populations underlies the brain’s processing of rich sensory information in our daily lives. Recent neuroscience and neuroimaging studies have increasingly used naturalistic stimuli and experimental design to identify such realistic sensory computation in the brain. However, existing methods for cross-areal interaction analysis with dimensionality reduction, such as reduced-rank regression and canonical correlation analysis, have limited applicability and interpretability in naturalistic settings because they usually do not appropriately ‘demix’ neural interactions into those associated with different types of task parameters or stimulus features (e.g., visual or audio). In this paper, we develop a new method for cross-areal interaction analysis that uses the rich task or stimulus parameters to reveal how and what types of information are shared by different neural populations. The proposed neural demixed shared component analysis combines existing dimensionality reduction methods with a practical neural network implementation of functional analysis of variance with latent variables, thereby efficiently demixing nonlinear effects of continuous and multimodal stimuli. We also propose a simplifying alternative under the assumptions of linear effects and unimodal stimuli. To demonstrate our methods, we analyzed two human neuroimaging datasets of participants watching naturalistic videos of movies and dance movements. The results demonstrate that our methods provide new insights into multi-regional interaction in the brain during naturalistic sensory inputs, which cannot be captured by conventional techniques.
\end{abstract}

\section{Introduction}

Naturalistic stimuli have been increasingly used in experimental neuroscience in both human \citep{Huth2012,Hasson2004} and animal studies \citep{Stringer2019} to investigate richer and more realistic stimulus-induced neural activities than those in traditional experimental design (Figure 1a). Recent advancements in data-driven feature extraction methods, including deep neural network \citep{Guclu2015, Mathis2018}, also contribute to this trend by drastically reducing the annotation costs for complex stimuli. One of the biggest challenges of using naturalistic stimuli is that they are usually non-categorical and high-dimensional, which is in stark contrast to traditional stimuli that are typically designed to be categorical and low-dimensional.

In addition to the increasing complexity of experimental stimuli, in recent neuroscience studies, there has been an increasing emphasis on investigating complex interactions between neural populations across brain areas \citep{van2013wu,simony2016dynamic,Kaufman2014,Semedo2019,Steinmetz2019} (Figure 1b) rather than simple stimulus-response relationships of individual neuron (or voxel) groups. This is important because the brain is a network system that consists of a tremendous number of neurons that interact with each other. The ongoing revolution in both measurement techniques and the scale of data provides neuroscientists with a fascinating opportunity to advance the understanding of cross-areal interactions in both neurophysiology \citep{Steinmetz2019} and human neuroimaging \citep{van2013wu,simony2016dynamic}. In particular, several authors have emphasized the importance of analyzing ‘multivariate’ (multi-to-multi) interactions between brain areas beyond ‘univariate’ interactions identified among individual neurons/voxels or their simple averages across areas. Several statistical techniques for dimensionality reduction have been used \citep{Cunningham2014} to explore multivariate cross-areal interactions in an interpretable, and computationally and statistically efficient manner. 

\begin{figure}
  \centering
  \includegraphics[width=\linewidth]{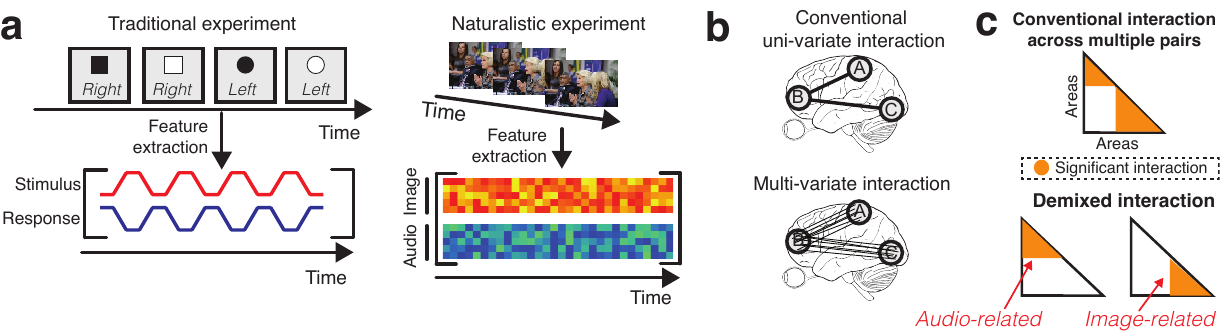}
  \caption{\textbf{Basic ideas and properties of the proposed NdSCA.} $\textbf{a}$ In traditional experiments (left), task parameters are usually pre-defined, categorical, and low-dimensional, whereas in naturalistic experiments (right), an arbitrary number of task parameters can, in principle, be extracted from continuous, high-dimensional, and usually multimodal (e.g., images or audio) naturalistic stimuli. $\textbf{b}$ Comparison among different types of interactions. Areas A, B, and C are communicating in a univariate (top) or multivariate  (bottom) manner. $\textbf{c}$ Mixed (top) and demixed (bottom) interactions in matrix format. Coloured areas indicate significant interaction. Only the lower-triangular part is visualized as we primarily assume symmetric interactions}
\end{figure}

However, the use of naturalistic stimuli complicates the analysis of cross-areal interactions, which poses several technical challenges. Existing methods that use canonical correlation analysis (CCA) \citep{Steinmetz2019} or reduced-rank regression (RRR) do not take external variables into account, which results in cross-areal connections that are ‘mixed’ in terms of their association with task parameters, such as stimulus features and other contextual factors (Figure 1c). The uncontrolled, high-dimensional, and multi-modal nature of naturalistic stimuli further makes the scenario extremely difficult because previous attempts in task-related neural interaction analysis \citep{friston1997psychophysiological,Rissman2004,Brendel2011,Kobak2016,Ito2017} commonly assumed traditional experimental designs and thus suffered from conceptual or computational problems. In fact, how to efficiently ‘demix’ cross-areal interactions into those associated with different task parameters of naturalistic experiments is an open problem. Novel technical advances are strongly desired so that one can fully take advantage of large-scale datasets and the rich information in naturalistic stimuli. 

In this paper, we propose neural demixed shared component analysis (NdSCA)\footnote{All code will be published after acceptance.}, as well as a simplified alternative, as a novel technique for cross-areal neural interaction analysis using dimensionality reduction. The method extends demixed shared component analysis (dSCA) \citep{Takagi2020}, proposed previously for traditional experimental design, in a non-trivial manner (see Table 1 for comparison with other methods). In particular, we propose using a recently proposed neural-network implementation of functional analysis of variance (ANOVA) with latent variables \citep{martens2020neural} to efficiently demix the complicated nonlinear effects of naturalistic stimuli into those associated with different types of task parameters. The continuous and multimodal nature of naturalistic stimuli cannot be handled by the conventional ANOVA on which dSCA relies. Our NdSCA combines this novel demixing technique with linear dimensionality reduction so that interactions among brain regions and their association with various task parameters are detected and visualized in an interpretable, and both statistically and computationally efficient manner. 

This paper is organized as follows. First, we briefly review previous methods for cross-areal interaction analysis, in addition to methods related to the idea of demixing (Section 2), and contrast their availability in traditional and naturalistic scenarios in neuroscience experiments. Then, we introduce the underlying ideas and specific algorithms of NdSCA and its simplified version (Section 3). Finally, we validate our methods using both simulation and applications to two human functional magnetic resonance imaging (fMRI) datasets (Section 4), and then conclude with a brief discussion (Section 5).

\begin{table}[]
\begin{tabular}{crcccccc}
\multicolumn{1}{l}{}         & \multicolumn{1}{l}{} & FC & PPI & RRR & dSCA & NdSCA-simple & NdSCA \\ \hline
\multirow{2}{*}{\begin{tabular}{c}Interaction\\among areas\end{tabular}} & demixed              & -  & \checkmark   & -   & \checkmark    & \checkmark            & \checkmark     \\
                             & multivariate       & -  & -   & \checkmark   & \checkmark    & \checkmark            & \checkmark     \\ \hline
\multirow{4}{*}{\begin{tabular}{c}Covariate\\(task parameter)\end{tabular}}   & multiple             & -  & \checkmark   & -   & -    & \checkmark            & \checkmark     \\
                             & multimodal          & -  & \checkmark   & -   & \checkmark    & -            & \checkmark     \\
                             & continuous           & -  & \checkmark   & -   & -    & \checkmark            & \checkmark     \\
                             & nonlinear    & -  & -   & -   & -    & -            & \checkmark     \\ \hline
\end{tabular}
\caption{\label{tab:table-name}Comparison among the methods for detecting interaction among areas.}
\end{table}

\section{Related work}
\label{gen_inst}
\paragraph{Detecting information sharing across different brain areas.}
In previous studies, information sharing among different brain areas was investigated for different scales from pairs of neurons \citep{Nowak1999} to interaction among multi-voxels \citep{Ito2017,Coutanche2013}. Particularly in human neuroimaging, a simple univariate correlation between large brain regions of interest (ROIs) is often calculated, which is called functional connectivity (FC). In line with the increased scale of datasets, both in the field of neurophysiology and human neuroimaging, researchers have started to investigate multivariate information sharing. Because simple regression often does not work in a high-dimensional setting, they have used statistical techniques such as principal component analysis (PCA) \citep{Kaufman2014,Ito2017}, RRR \citep{Semedo2019}, or CCA \citep{Steinmetz2019} to effectively reduce the dimensionality. However, these methods do not provide results that are demixed with respect to experimental task parameters. This is problematic because, although the results of several studies have suggested that task-related information sharing might have richer information than that of the resting state \citep{finn2021movie}, the content of the shared information is unclear.

\paragraph{Demixing task parameters}
It is well known that even a single neuron has mixed selectivity, i.e., the neuron’s firing rate responds to more than one task parameter \citep{Rigotti2013}. This is true even after dimensionality reduction \citep{Mante2013}. Mixed selectivity is also commonly observed in a large-scale voxel in human fMRI \citep{Huth2012}. It is a critical problem for researchers who are interested in cognitive processing during complex experiments because the analysis and interpretation then become complicated in order to dissociate different types of computations run simultaneously in the brain. To overcome this problem, several approaches have been proposed. Among others, demixed PCA (dPCA) \citep{Brendel2011, Kobak2016} has successfully been used in neurophysiology, which was recently extended to dSCA to capture information sharing  among areas \citep{Takagi2020}. The key idea is to ‘marginalize’ multivariate information in a single area in association with a specific task parameter of interest. These methods allow experimenters to combine rich data with equally rich task design, identifying low-dimensional components that vary along axes defined by features of the experimental task. In human neuroimaging, psychophysiological interaction (PPI) \citep{friston1997psychophysiological} and beta series correlation (BSC) \citep{Rissman2004}, which are closely related mathematically \citep{Di2021}, have specifically been used for a long time but they are designed for traditional experiments. In fact, PPI and BSC primary consider univariate interactions and also not applicable to large-dimensional task parameters with a reasonable computational cost.

\section{Neural demixed shared component analysis (NdSCA)}
\subsection{Cross-areal interaction analysis with dimensionality reduction}

Our goal is to investigate task or stimulus-related interactions among brain areas induced by naturalistic stimuli in an efficient and interpretable manner. Assume that brain activity is measured simultaneously in two areas. For each area, we observe the multivariate time series of firing rates or blood oxygen level-dependent signals sampled at T discrete time points. We then provide a conventional FC measure using the correlation of average activities between two areas, but this is suboptimal if the interaction is not through averages of the areas. 

Linear dimensionality reduction techniques are particularly useful for evaluating the multivariate interaction of neuron/voxel groups between brain areas. A popular method in the neuroscience literature is RRR. Let $\mathbf{X}$ and $\mathbf{Y}$ denote given data matrices of two areas, with sizes $M_{X} \times T$ and $M_{Y} \times T$, respectively, where $M_{X}$ and $M_{Y}$ denote the number of neurons or voxels in respective areas and T denotes the time length. Then, standard least-squares RRR minimizes the following:
\begin{equation}\label{eq:RRR}
  L_\mathrm{RRR} = \left \|   \mathbf{Y} -  \mathbf{WX}   \right \|^{2},
\end{equation}
where $\|.\|$ represents the Frobenius norm; $\mathbf{W}$ is a coefficient matrix of size $M_{X} \times M_{Y}$ and its rank is constrained to not be greater than $P$ $(<\min(M_X,M_Y))$. The solution $\mathbf{W}_{RRR}$ is obtained as
\begin{equation}\label{eq:RRRsolution}
    \mathbf{W}_{RRR} = \mathbf{W}_{OLS}\mathbf{VV}^T,
\end{equation}
where $\mathbf{W}_{OLS}$ is the ordinary least-squares solution and the columns of the $M_{X} \times P$ matrix $\mathbf{V}$ contain the top $P$ principal components of the optimal linear predictor $\mathbf{\hat{Y}}_{OLS} := \mathbf{W}_{OLS}\mathbf{X}$ \citep{Kobak2016}. Note that RRR reduces to CCA if $\mathbf{Y}$ is whitened  \citep{Torre2012}. Given an RRR solution, we define the interaction strength between the two areas as the total explained variance (EV) of the target activity.
 
Although RRR might achieve higher sensitivity than simple correlation, RRR still fails to capture task-related information sharing, that is, results are mixed in terms of task parameters. Therefore, in the following, we consider task-related information sharing explicitly.

\subsection{Demixing effects of naturalistic stimuli via functional ANOVA}

Suppose that we are given multiple task parameters of interest, denoted by $c_k$ ($k=1,2,\ldots,K$),  where $K$ is the number of different types of task parameters, in addition to brain activity measurements. First, we consider traditional experiments such that $c_k$ are discrete (univariate) and K is relatively small. Then, their induced effects on brain activity may be modeled using the well-known framework of (multivariate) ANOVA, which underlies the theory of dPCA and its extension for cross-area interaction analysis, that is, dSCA. The key idea is that ANOVA eventually decomposes a given data matrix $\mathbf{Y}$ into multiple matrices, each corresponding to the main or interaction effects by the discrete task parameters; for instance, if $K = 2$, two-way ANOVA identifies the decomposition, given by
\begin{equation}
    \label{Eq:anova}
    \mathbf{Y} = \mathbf{Y}_1 + \mathbf{Y}_2 + \mathbf{Y}_{12}  + \mathbf{E},
\end{equation}
where $\mathbf{Y}_0$ corresponds to the overall mean; $\mathbf{Y}_1$ and $\mathbf{Y}_2$ correspond to the main effects of $c_1$ and $c_2$, respectively; $\mathbf{Y}_{12}$ corresponds to the interaction effects; and $\mathbf{E}$ corresponds to residuals that are not explained by task parameters.\footnote{Note that in each matrix, except for $\mathbf{E}$, the columns are identical if their associated task-parameter values are the same; for example, $\mathbf{Y}_0$ replicates the overall mean vector across columns, and the columns in $\mathbf{Y}_1$ with the same value of $c_1$ are identical, each given by the corresponding $c_1$-conditional empirical mean.} The procedure to obtain the ANOVA decomposition in addition to each resultant term was called marginalization in \citep{Kobak2016,Takagi2020}. Given the decomposition, dSCA between source $\mathbf{X}$ and target $\mathbf{Y}$ is formalized as solving RRR for each marginalization of $\mathbf{Y}$, with the $\mathbf{Y}$ in Eq.\eqref{Eq:anova} replaced by the corresponding term in the ANOVA decomposition; dSCA reduces to dPCA if the source and target areas are the same. Intuitively, each RRR then identifies a rank-reduced connection from the source to target areas such that the connection specifically explains the target variability related to the particular subset of task parameters.

Generalizing the idea of dSCA in naturalistic settings is, however, not straightforward. The problem is that the idea of marginalization strongly relies on conventional ANOVA, and thus cannot directly be applicable if a task parameter $c_k$ is continuous. Moreover, features obtained from naturalistic stimuli are usually rather high-dimensional, even in each individual modality, such as visual or audio. In practice, modeling the effect of high-dimensional task parameters easily leads to high-computational complexity if not combined with any suitable parametric architecture and associated algorithms. Another potential issue is that all the task parameters of interest would not necessarily be observed (extracted), which is more likely to occur in naturalistic experiments than traditional experiments as the stimuli are not strongly controlled.

To overcome all these issues, we propose extending the basic idea of dSCA using neural decomposition (ND)  \citep{martens2020neural}, which is a recently proposed practical instantiation of functional ANOVA \citep{sobol2001global} with a conditional variational autoencoder (CVAE) \citep{Kingma2014}. Functional ANOVA generalizes the marginalization of dSCA in a principled manner, and the use of the deep neural network technique enhances the practical utility in high-dimensional settings. We refer to the resultant analysis framework as NdSCA. In practice, the algorithm simply replaces the marginalization part of dSCA with the generalized part using ND (Algorithm 1). Therefore, in the following, we focus on presenting the basic idea of ND and its interpretation from a naturalistic experiment viewpoint.
\begin{figure}[!t]
\begin{algorithm}[H]
    \caption{Neural demixed shared component analysis (NdSCA) between two brain areas}
    \begin{algorithmic}[1]
    \REQUIRE Matrices $(\mathbf{X},\mathbf{Y})$ of source and target brain activity; those of $K$ task parameter vectors $\mathbf{c}_k$
    \ENSURE Rank-reduced coefficient matrices $\mathbf{W}_{RRR}$ for all relevant marginalizations 
    \STATE Run ND \citep{martens2020neural} to decompose $\mathbf{Y}$ into marginalizations $\mathbf{Y}_{\phi}$ corresponding to a functional ANOVA
    \FOR{each relevant marginalization $\mathbf{Y}_{\phi}$ (e.g. $\phi \in \{z,c,zc\}$ for the model \eqref{Eq:fanova})}
    \STATE Solve an RRR problem \eqref{eq:RRR} from $\mathbf{X}$ to $\mathbf{Y}_{\phi}$ with Eq.\eqref{eq:RRRsolution}
    \ENDFOR
    \end{algorithmic}
\end{algorithm}
\end{figure}

Now, we consider a naturalistic setting in which each task parameter, denoted by ${\bf c}_k$, may be continuous and vector-valued. For example, ${\bf c}_1$ and ${\bf c}_2$ represent visual and audio features of the naturalistic stimulus extracted at each time frame of the brain activity measured. To model the potential effects from unobserved but relevant task parameters, which are more likely to exist in naturalistic settings than traditional settings, we further introduce an additional random vector z of latent variables. To ease exposition, assume for simplicity that the number of observed task-parameter vectors is $K = 1$. Then, we can generally write the task-induced effect on brain activity $\mathbf{y}$ as $F({\bf z,c})$, where $F$ denotes any appropriate nonlinear mapping. The idea of functional ANOVA and thus ND is to decompose $F({\bf z,c})$ into marginal (main) and interaction effects of ${\bf c}$ and ${\bf z}$, such that
\begin{equation}
    \label{Eq:fanova}
    \mathit{F}(\bf{z,c}) = \mathit{F}_\mathrm{0} + \mathit{F}_\mathrm{z}(z) + \mathit{F}_\mathrm{c}(c) + \mathit{F}_{\mathrm{zc}}(z,c)
\end{equation}
where $F_0$ denotes the intercept and other terms; $F_c(\mathbf{c})$ and $F_z(\mathbf{z})$ represent the main effects of ${\bf c}$ and ${\bf z}$, respectively; and $F_{\mathrm zc}(\bf z, c)$ represents their interaction effect. Generalization with more than one task-parameter vector $\mathbf{c}_k$ is straightforward, with their main and interaction effects appropriately included up to a reasonable order. Note that a decomposition such as Eq.\eqref{Eq:fanova} is clearly non-unique if not further constrained; functional ANOVA ensures the identifiability of the model by introducing constraints on each term such that its integration over every input argument is zero \citep{martens2020neural}, which naturally generalizes the typical sum-to-zero constraint in classical ANOVA.

Although the non-parametric form of functional ANOVA above is theoretically relevant, it is not practically useful and thus needs to be further implemented with parametric architectures, which must easily incorporate possibly high-dimensional $\bf{c}_k$ and latent input $\bf z$ in our context. From this viewpoint, the recent neural network-based realization, ND, of functional ANOVA is particularly suitable as it offers flexible and computationally efficient implementation based on a CVAE \citep{Sohn2015}, explicitly modeling and performing inference on latent variables $\bf z$. Briefly, a CVAE learns a parametric generative model $p_{\theta}(\bf{y}|\bf{z,c})$ defined by the decoder network $\bf{y} = F_{\theta}(\bf{z,c})$ and a prior $p(\bf{z})$, while approximating the posterior $p(\bf{z}|\bf{y,c})$ parametrically with an encoder network $q_{\phi}(\bf{z}|\bf{y})$. Both the parameters $\phi$ and $\theta$ are trained by approximately maximizing the variational lower bound of the conditional log-likelihood (i.e., $\log p_{\theta}(\bf{y}|\bf{c})$ for a single datum), typically using stochastic gradient ascent. In ND, the additive form and specific input dependences of each term in Eq.\eqref{Eq:fanova} are explicitly modeled in the decoder network, and penalty and augmented Lagrangian methods are developed to deal with the integral constraints (see \citep{martens2020neural} for details). 

Given the decoder network learned, we can obtain an additive matrix decomposition similar to Eq.\eqref{Eq:fanova} by first estimating latent vector $\bf z $ for every column of the target matrix $\mathbf{Y}$ and then computing terms such as $F_c(\mathbf{z,c})$ at each instance of $\bf z $ and $\bf c$. Then, if we perform RRR as above by regarding each of these terms as a generalized marginalization of $\mathbf{Y}$, NdSCA obtains rank-reduced cross-areal interaction and low-dimensional components of activities in the areas. We quantify the interaction so that it is sensitive only to a particular (main or interaction) effect of observed and unobserved task parameters. Note that the final RRR step of NdSCA provides a linear relationship among voxels/neurons and thus the contribution of each voxel/neuron can be easily visualized and interpreted. This interpretability is crucial for neuroscientists; previous approaches based on decoding \citep{Coutanche2013} or representational similarity analysis \citep{Ito2017} did not provide it.

\subsection{Simplified alternative}

The novel marginalization procedure of NdSCA can be applied to various scenarios of neuroscientific data analysis, even in non-naturalistic settings, because of the flexibility of ND and the underlying functional ANOVA model. However, the method can be overly complicated if we consider more restricted and rather controlled (more traditional) scenarios in which the effect of non-observed task parameters (latent variables $\bf z$) can be ignored. In particular, a simple alternative would be sufficient if only a single type (modality) of continuous stimuli was used and the linearity of stimulus effects could be reasonably assumed. 

Specifically, we consider a single multivariate task parameter, which is continuous and possibly high-dimensional. Let $\mathbf{C}$ denote the corresponding data matrix, with the instances of the task parameter vectors as its columns. Then, if no latent variables are necessarily introduced, the linear effect on a single-region activity is modeled simply by $\mathbf{Y} = \mathbf{WC} + \mathbf{E}$ without resorting to more general functional ANOVA modeling. We can therefore use multivariate linear regression or similar techniques to identify the task-specific effect in $\bf{Y}$, that is, $\bf{Y}_{\bf{C}}:=\bf{WC}$; in this case, we specifically use the same RRR/CCA technique as above to improve the identification.

In fact, this simple alternative, which we refer to as NdSCA-simple, is novel and useful in its own right. This also offers a reasonable baseline for validating the performance of NdSCA in our simulation and real-data analysis below. Note, however, that the method is suboptimal if the simplifying assumptions are violated, for example, when we need to simultaneously incorporate multiple types (modalities) of task parameter vectors, as is typical in naturalistic experiments.

\section{Results}
\paragraph{Synthetic data}
To confirm that demixing is necessary for capturing task-parameter-specific information sharing and NdSCA can deal with naturalistic experiments for that purpose, we generated simulated neuronal populations that is analogous to the naturalistic experiment.

Suppose that we simultaneously recorded population neural activities in three brain areas, A, B, and C, during a traditional experiment with two task parameters of interest: Audio and Image (see Figure 1a right for an illustration). Both parameters are high-dimensional and continuous, such as features extracted using an artificial neural network. We set the dimension of the task parameters and neurons to 10 and 40, respectively. We synthesized the data so that the interactions (co-activations) of neural activity between areas A, B, and C varied in response to Audio and those between A and C varied in response to Image, as indicated by the two colors in Figure 2a.  We simulated $3 \times 10^4$  observations, and split them into training, validation, and test data (1000/1000/1000). See the Supplementary Material for the details.

For comparison, we applied FC, RRR, PPI, and our NdSCA to the dataset. For each method, we calculated the EVs of predictions from one region to another. To calculate the strength of interactions, for example, between areas A and B, we averaged the EVs of predictions from A to B and B to A. We calculated the statistical significance of interactions by running permutation testing 1,000 times. For PPI and NdSCA, we also calculated the statistical significance for each area in terms of the representation of the specific task parameter using ND, and calculated the strength of interactions only if both source and target areas had information about the task parameter. To train ND, we used the Adam optimizer \citep{Kingma2014}. We trained the model at a maximum of 50 epochs with a batch size of 128. We used early stopping with a patience of 10 epochs. We set the number of latent units $\bf z$ to 1. We used a warm-up process that weighted the reconstruction loss more than the Kullback–Leibler loss \citep{pmlr-v48-maaloe16}. For FC and PPI, we applied PCA to reduce the dimensions of neurons to one, whereas for RRR and NdSCA, we used multi-dimensional data without PCA. We also applied PCA to the task parameters to reduce the dimensions to three before applying PPI and NdSCA.

\begin{figure}
  \centering
  \includegraphics[width=\linewidth]{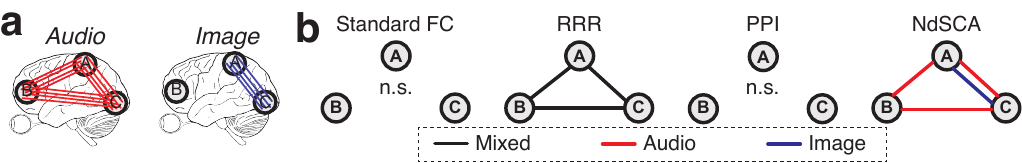}
  \caption{\textbf{Simulations demonstrate that only NdSCA detected complex task-parameter specific interaction in the naturalistic experiment.} $\textbf{a}$ Areas A, B, and C communicate Audio information (top; indicated by red lines), whereas areas A and C communicate Image information (bottom; indicated by blue line). $\textbf{b}$ Results obtained from standard FC, RRR, PPI, and NdSCA ordered from left to right. Only NdSCA detected demixed interaction. n.s. corresponds to a no significant interaction.}
\end{figure}

We found that FC and PPI could not detect any information sharing because they do not use an underlying multivariate structure (Figure 2b; $P > 0.05$). Although RRR detected information sharing among areas ($P < 0.05$ for lines in Figure 2b), it did not indicate what information was shared across areas. Strikingly, only NdSCA revealed the demixed underlying information sharing structure ($P < 0.05$ for coloured lines in Figure 2b). This is because only NdSCA could manage multivariate information with complex task parameters. 

\paragraph{Naturalistic experiment – movie watching fMRI}
Next, we applied NdSCA to a human fMRI dataset of participants watching movies \citep{Aliko2020} to validate the advantage of our methods. The dataset consisted of 86 participants watching 10 full-length movies. We used six participants watching ‘The Shawshank Redemption’ that lasted for 8,181 s (Figure 3a). The authors also provided time-aligned Word (transcribed from audio) and Face (appear/not appear) annotations for each movie. We used a binary feature vector for Face and three-dimensional latent semantic matrices for Word that were converted using fastText (\citep{Bojanowski2017}; see the Supplementary Material). We split all scans into training, validation, and test data (with 4,800, 1,200 and 2,181 scans, respectively).

We investigated information sharing among small-scale ROIs and also among large-scale functional networks. Specifically, we used 10 ROIs obtained from automated anatomical labeling (\citep{Tzourio-Mazoyer2002}; Figure 3b, see Supplementary Material), which are related to faces, words and associative functions, and eight functional networks in \citep{Finn2015} (Figure 3e). Before applying ND, we applied PCA to each ROI/network to reduce the dimensionality to either 20 (as source) or 50 (as taarget). 
NdSCA was run based on a functional ANOVA model for the two types of task parameters (i.e. face and word) and one latent vector $\mathbf{z}$, while their interaction terms were not included to simplify the analysis. We calculated EVs for each participant, and averaged across participants for visualization and statistical testing.

Figures 3d and 3g show that RRR detected information sharing among almost all ROIs and networks ($P < 0.05$ for coloured pairs in Figures 3d and 3g; FWE corrected). However, as shown in the simulation analysis, they did not indicate what information they shared. While PPI detected task-related interaction to some extent, e.g., Face-related interaction between fusiform face area (FFA) and V1, PPI failed to detect Word-related information sharing. Strikingly, only NdSCA successfully revealed Word- and Face-related information sharing. The ROIs and networks that were known for visual information processing (e.g., FFA for the ROIs; Visual 1 and Visual 2 networks) exhibited Face-related information sharing with each other, whereas areas known for audio and semantics (e.g., A1 and temporal areas for the ROIs; medial temporal/frontoparietal networks) Word-related information sharing ($P < 0.05$ for coloured pairs in Figures 3d and 3g; FWE corrected). Notably, NdSCA had much higher sensitivity than NdSCA-simple, which might suggest that the underlying relationship between brain activity and the task parameters is nonlinear.

We also investigated the demixed interaction associated with the effect of latent variables $\mathbf{z}$ (representing non-observed but relevant task or stimulus parameters) by applying NdSCA with the $F_z(\mathbf{z})$ term in Eq.\eqref{Eq:fanova} reconstructed for every estimated instance of z. The result appeared to be the same as that obtained by non-demixed RRR (Supplementary Figure 1), although we did not investigate their differences in detail as this is beyond the scope of this paper. The apparent similarity, at least, indicates that a large number of variations in cross-regional interaction can be nonlinearly explained by a relatively small number of factors that are not related to Face or Word. This is in fact reasonable because naturalistic movies contain many other features and also cross-areal interaction is strongly affected by the intrinsic interaction, reflecting spontaneous brain activity, even during tasks \citep{Tavor2016, Cole2016}.

\begin{figure}
  \centering
  \includegraphics[width=\linewidth]{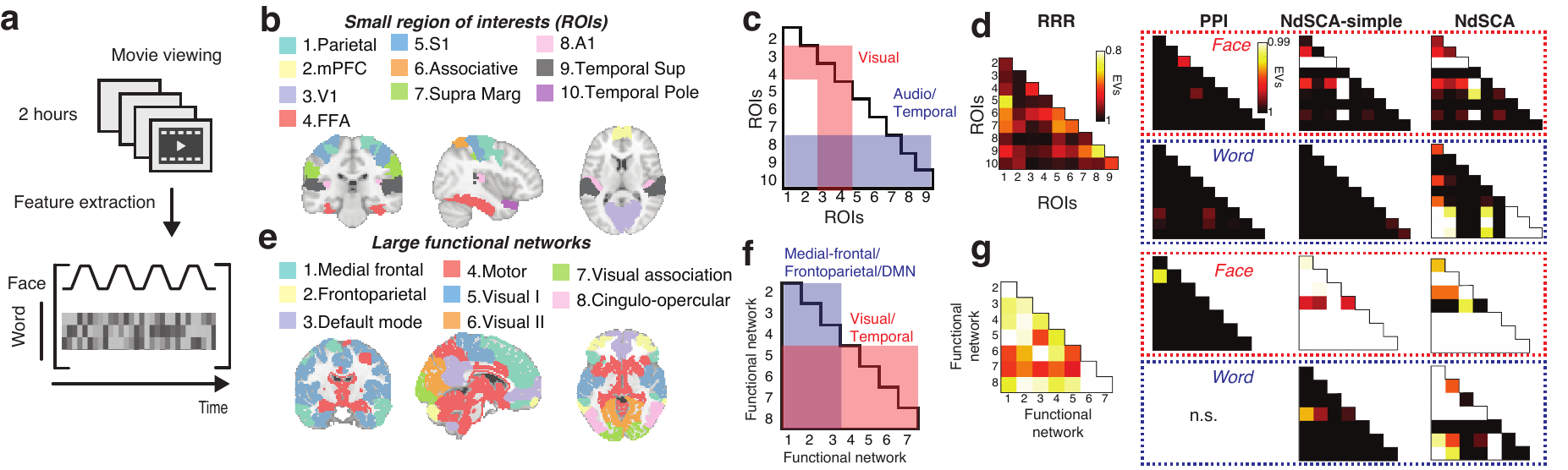}
  \caption{\textbf{NdSCA successfully demixed different types of computation.} $\textbf{a}$ Schematic of the task. Human participants watched 2-hours of a movie. We extracted a binary face vector (appear/not appear) and multi-dimensional latent semantic matrices as features. $\textbf{b}$ We used 10 ROIs. See the Supplementary Material. $\textbf{c}$ ROIs were divided into visual, audio/temporal, and others. $\textbf{d}$ Results obtained from RRR, PPI, NdSCA-simple, and NdSCA are ordered from left to right. For PPI and NdSCAs, the results of Face (top) and Word (bottom) are displayed separately. $\textbf{e-g}$ We also conducted similar analysis for eight functional networks. Non-significant pairs are coloured black. n.s. corresponds to a no significant interaction.}
\end{figure}

\paragraph{Naturalistic experiment – Dance clip fMRI}
Finally, we applied NdSCA to a human fMRI dataset of participants watching dance clips (Figure 4a). Participants watched 1,163 dance clips that were performed by 30 dancers from 10 genres for 10–50 seconds \citep{tsuchida2019aist,li2021learn}, which resulted in nearly five hours and approximately 15,000 scans (see the Supplementary Material).\footnote{We collected the fMRI data used in this experiment.} We split all scans into training, validation, and test data (with 10,874, 1,200 and 2,756 scans). Note that the dance clips in the test dataset consisted of music and choreography that did not appear in both training and validation datasets. We used the same ROIs and networks as in the previous analysis. 

We used automatically extracted audio and motion features from the clips. For audio features, we used the publicly available audio processing toolbox Librosa \citep{mcfee2015librosa} to extract a one-dimensional envelope, mel-frequency cepstral coefficients, and chroma, resulting in a 33-dimensional feature vector. For the motion features, we used 17 COCO-format three-dimensional (3D) human joint locations, which resulted in a 51-dimensional feature vector \citep{li2021learn}. We applied PCA to reduce the dimensionality of each feature to three. Again, NdSCA with the two type of feature (task parameter vectors) and one latent vector was used with no interaction terms among them for simplicity. See the Supplementary Material for the details.

Figures 4c and 4e show that, again, only NdSCA successfully detected task-parameter-specific information sharing ($P < 0.05$ for coloured pairs in Figures 4c and 4e; FWE corrected). It shows that audio-related information sharing was propagated across the brain, whereas motion-related information sharing was more strongly observed around the visual ROIs/networks.

\begin{figure}
  \centering
  \includegraphics[width=\linewidth]{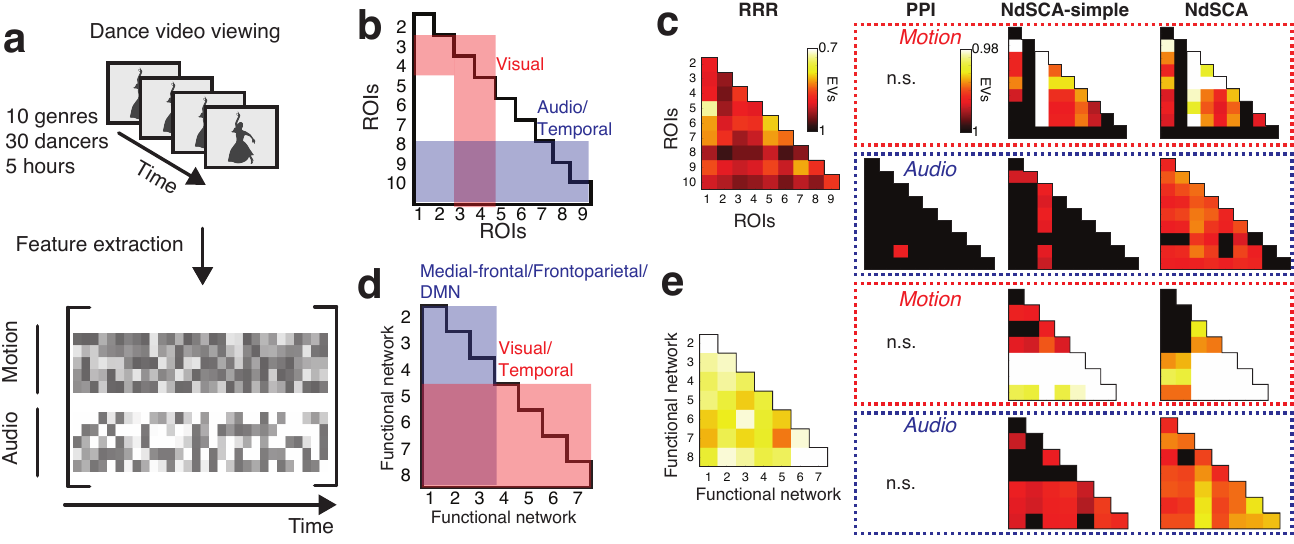}
  \caption{\textbf{NdSCA demixed information sharing into meaningful task parameters, whereas the conventional method failed.} $\textbf{a}$ Schematic of the task. Human participants watched 5-hours of dance clips. We extracted multi-dimensional 3D joint locations (motion) and audio features. $\textbf{b}$ We used the same ROIs as in the previous analysis. $\textbf{c}$ Results obtained from RRR, PPI, NdSCA-simple, and NdSCA are ordered from left to right. $\textbf{d-e}$ Same analyses for eight functional networks. Non-significant pairs are coloured black. n.s. corresponds to a no significant interaction.}
\end{figure}

Given that only NdSCA provided task-specific information sharing, we further investigated whether NdSCA could also reveal the different information sharing patterns between different conditions. Although previous studies investigated different patterns of interaction between different conditions \citep{simony2016dynamic}, there measures are mixed in terms of task-parameters, thus make it difficult to interpret the results. In this dataset, there was a control condition under which participants watched dance clips played backward (Fig 5a). We hypothesized that pattern of interaction among areas in the backward-play condition was substantially changed from those in the forward-play condition. We also hypothesized that the effects were not equal across different task parameters, audio and motion.

\begin{figure}
  \centering
  \includegraphics[width=\linewidth]{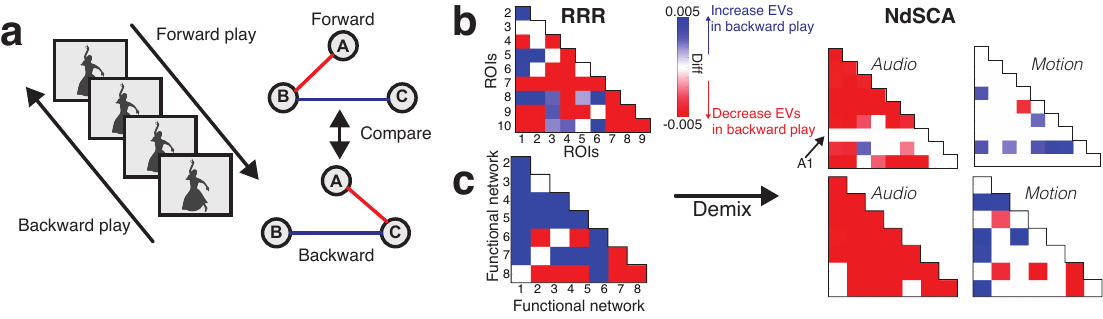}
  \caption{\textbf{NdSCA reveals strong double-dissociation of effects of backward playing between audio- and motion-related information shraing} $\textbf{a}$ Comparison between forward and backward play. $\textbf{b-c}$ NdSCA (right columns) indicated that interactions for audio and motion were differently manipulated by backward playing, whereas RRR could not demonstrate this (left column) for the ROIs ($\bf b$) and networks ($\bf c$). Non-significant pairs are coloured white.}
\end{figure}

Figures 5b and 5c show that NdSCA confirmed our hypothesis: manipulation effects on audio and motion-related information sharing were drastically different, rather than demonstrating a mere overall decrease/increase of information sharing ($P < 0.05$ for coloured pairs in Figures 5b and 5c; FWE corrected). Specifically, audio-related information sharing changed globally, whereas motion-related information sharing did not change consistently across the ROIs/networks. Interestingly, audio-related interaction around A1 did not suffer even in the condition of backward playing (indicated by the black arrow). All the results above could never be observed by conventional connectivity methods, including RRR, which suggests that our proposed method had an advantage over existing methods \citep{simony2016dynamic,finn2021movie}.

\section{Conclusion and discussion}
Here we proposed NdSCA, a new technique for demixing complex task parameter specific information sharing among brain areas during naturalistic experiments. With both simulations and naturalistic fMRI data analyses, we demonstrated that NdSCA can provide a novel neuroscience insight by demixing patterns of cross-areal information sharing into those associated with different types of task or stimulus parameters (features). Previously, non-demixed types of techniques have commonly been used in cross-areal interaction analysis and thus often yielded results that were not easily interpreted.

However, our method has several limitations. First, in return for an expression power, NdSCA needs relatively larger data compared to simple linear methods. However, given that the size of the publicly available dataset is getting larger, we believe that NdSCA can contribute to the trend. Second, the current implementation of NdSCA only captures linear information sharing among areas. Although it provides interpretability, future study should consider non-linear information sharing among areas. 

Future research topics include: (i) applying NdSCA to time-resolved data obtained by, e.g. microelectrodes or magnetoencephalography; (ii) investigating the behavioural relevance of the shared component; (iii) extending NdSCA to deal with more than two areas, that can be easily implemented thanks to the flexible architecture of NdSCA.

\begin{ack}
YT was supported by JSPS overseas fellowship. LH was supported by a Sir Henry Dale Fellowship from Wellcome and the Royal Society (208789/Z/17/Z). HI was supported by JSPS KAKENHI (19H05725 and 21H00959) and AMED (JP18dm0307008). JH was supported by a project, JPNP20006, commissioned by the New Energy and Industrial Technology Development Organization (NEDO), and by JSPS KAKENHI (18KK0284 and  21K12055). 
\end{ack}

\bibliography{neurips_2021}

\end{document}


\maketitle

\section{Neural dSCA}
\subsection{Model architecture of Neural Decomposition (ND)}

We used a common neural architecture for all the three experiments in Section 4. Specifically, ND had one encoder, $F_{\bf{Y}}(\mathbf{Y},\mathbf{c}_1,...,\mathbf{c}_K)$, where $\mathbf{Y}$ is a given data matrix of the target area with size $M_{Y} \times T$, $M_{Y}$ denote the number of neurons or voxels in the area, $T$ denotes the time length, $K$ is the number of different types of task parameters, and $\mathbf{c}_k$ ($k=1,2,\ldots,K$) denotes the matrix of k-th task parameter, $F$ denotes nonlinear mapping; one decoder for the latent variable $\bf z$, $F_{\bf{z}}(\bf{z})$; and multiple decoders that corresponded to each covariate $\mathbf{c}_1,..., \mathbf{c}_K, F_{c_{1}}(\mathbf{c}_{1}),..., F_{c_K}(\mathbf{c}_K)$. Because all the datasets in the study had two covariates, we considered the case of K = 2. We also considered a 1D latent variable $\bf z$ throughout, although our approach was not restricted to a univariate $\bf{z}$. The encoder consisted of three linear fully connected layers ($M_Y + M_{c_{1}} + M_{c_{2}} \times 12,  12 \times 24, 24 \times 2M_z$). The output neurons of the third layer corresponded to the mean and s.t.d. of the latent variable $\bf z$, and we approximated the posterior distribution of the latent variable $\bf z$ using a multivariate Gaussian that had a diagonal covariance structure, zero mean, and unit standard deviation. We used ReLU activations throughout. The decoder $F_{c_{1}}$ and $F_{c_{2}}$ consisted of four linear fully connected layers ($M_{c_{k}} \times 24$, $24 \times 48$, $48 \times 24$, and $24 \times M_Y$), where $k = 1$ or 2 and $M_k$ denotes the dimension of $c_k$. We used tanh activations throughout. We did not include the interaction of covariates, $F_{c_{1}c_{2}}$, to focus on the main effect of each covariate.

\subsection{Training procedure of ND}
We set the batch size to 128 and trained ND for 50 epochs. We used the Adam optimizer with 0.0005 learning rate. We also used early stopping with a patience of 10 epochs and a maximum of 50 epochs. We used a warm-up schedule on the regularization objective by increasing the weight on the KL-loss from $10e-8$ linearly during training \citep{pmlr-v48-maaloe16}. To obtain an identifiable neural decomposition, we used the modified differential multiplier method that was proposed by the previous study \citep{martens2020neural} with a fixed constant c of 0.01. We estimated the integrals using the quadrature of 15 equally spaced points between [-6, 6] for an augmented Lagrangian. We trained and tested our model on one Tesla V100 GPU. The time required to train one region was approximately 5 to 10 minutes.

\subsection{Estimate a strength of interaction between areas}
After training ND, for each task parameter, we calculated the marginalized target matrix both on the training and testing dataset. Next, we applied RRR to the source and target matrices in the training dataset, with L2 regularized using $\lambda = 5$. We predicted the values of the marginalized target matrix from the source matrix in the test set using a transformation matrix estimated by the training set. We then computed the explained variance between the predicted and actual target matrices in the testing dataset. We repeated this procedure across all ROIs or functional networks. For NdSCA-simple, we applied CCA to the target matrix in the training dataset for each task parameter, and then applied the transformation matrix to the testing dataset. We then conducted the same procedure as above. 

\subsection{Statistical significance}
To calculate the statistical significance, we used 1,000 times permutation testing procedure. Specifically, in ND, to determine whether a given area A had information about task parameter $c_k$, we compared the reconstruction loss obtained from  $F_{c_{k}}$ in terms of $c_1,...,c_K$ between the true and randomly permuted $c_k$. To determine whether a connectivity strength is significant, we compared the explained variance obtained from RRR using true source matrix with the explained variances obtained from randomly permuted source matrices. 

\section{Simulation analyses}
To simulate a naturalistic experiment (Fig. 1a right), we simulated $3*10^4$ observations in three areas: A, B, and C. We split the dataset into training, validation, and test data (1000/1000/1000). Neurons in areas A, B, and C were communicating Stimulus information, and A and C were communicating Response information. Specifically, neurons in area A were affected by Stimulus information, and then passed it to neurons in area B. Neurons in area B further passed the information to neurons in area C. Finally, neurons in area C passed the information back to area A. Neurons in areas A and C were also communicating Response information with each other. These information transmissions were conducted via random projection matrices. We drew baseline activities for areas A, B, and C randomly from the Gaussian distribution $\mathcal{N}(0, 1)$. We drew two baseline task parameters, Stimulus and Response, from the Gaussian distribution $\mathcal{N}(0, 1)$. We set the dimensions of neural activities and task parameters to 40 and 10, respectively. We applied FC, RRR, PPI, and NdSCA, and tested the statistical significance as described above. We ignored variation across time in the current study, though we can apply our methods to time-resolved dataset. 

\section{Cinema movie dataset (Aliko et al., 2020)}
The authors provided a dataset that consisted of 86 participants watching 10 full-length movies from 10 genres. We used data from six participants (age 19–30; three females) watching “The Shawshank Redemption” that lasted for 8,181 sec. Each movie was watched with 40–50 minute intervals or when the participants asked for a break, which resulted in two to six runs. The data acquisition and preprocessing details, which include motion and physiological nuisance regression steps, are described in \citep{Aliko2020}. We obtained the reprocessed dataset and annotations from https://openneuro.org/datasets/ds002837/versions/2.0.0 and https://www.naturalistic-neuroimaging-database.org/, respectively. 

fMRI data were collected using a 1.5-T Siemens MAGNETOM Avanto with a 32- channel head coil (Siemens Healthcare, Erlangen, Germany). The authors used multiband EPI (TR, 1000 ms; TE, 54.8 ms; flip angle, 75 deg; number of slices, 40; voxel size, $3.2 \times 3.2 \times 3.2$ mm). A 10-minute high-resolution defaced T1-weighted anatomical MRI scan (MPRAGE) is also provided.

We extracted the time series of BOLD signals from each of 10 regions of interest (ROIs) or eight large-scale functional networks that were defined in a previous study \citep{Finn2015}. We obtained 10 ROIs from AAL \citep{Tzourio-Mazoyer2002} (Precentral gyrus [Parieal]; Frontal Superior frontal gyrus, medial [mPFC]; Calcarine [V1]; Fusiform gyrus [FFA]; Postcentral gyrus [S1]; Superior parietal gyrus [Associaive]; Supramarginal gyrus [Supra Marg]; Heschl [A1]; Superior temporal gyrus [Temporal Sup]; and Temporal Pole, Superior [Temporal Pole]). Because all the ROIs had right and left parts, we analyzed the right and left ROIs separately, and then averaged the results.

The authors also provided automated word and face annotations for each movie. Because the authors conducted annotation automatically, word annotation had four accuracy levels, “Matched,” “Continuous,” “Partial,” and “Full,” ordered from highest to lowest accuracy. We used the “Matched” or “Continuous” annotations for the analysis. To use stimulus annotations, timing correction was done to account for delays caused by the movie pausing script to assure that the fMRI time series and movies were well aligned. To convert each word into the latent semantic feature, we used fastText with a dimensionality of three \citep{Bojanowski2017}. For the Face vector and each dimension of the latent semantic feature, we averaged across at a 1-s resolution (i.e., BOLD sampling rate) and applied z-score normalization.

We applied RRR, PPI, NDSCA-simple, and NDSCA, and tested the statistical significance as described above, except for the decoder of Face. We used a simplified version of the decoder for $F_{C_{Face}}(C_{Face})$ that consisted of two fully connected layers ($1 \times 12$ and $12 \times M_Y$) because Face was a binary feature vector.

\section{Dance movie dataset}
We recorded whole-brain fMRI signals from six healthy participants (age 21–32; two females) while they were freely viewing 5-h dance movies that consisted of 1,163 clips. We selected dance clips from AIST Dance DB \citep{tsuchida2019aist} with AIST++ Dance Motion Dataset \citep{li2021learn}, the largest and richest existing dataset that contains 10 dance genres with 3D human keypoint annotation: Old School (Break, Pop, Lock, and Waack) and New School (Middle Hip-hop, LA-style Hip-hop, House, Krump, Street Jazz, and Ballet Jazz). We obtained datasets from https://aistdancedb.ongaaccel.jp/, and we obtained 3D motion from https://google.github.io/aistplusplus\_dataset/. The participants were instructed to watch the clips naturally as if watching dance/music clips in daily life. All the participants provided written informed consent, and none of the participants had risk factors associated with fMRI scanning (metal implants, claustrophobia, pregnancy, experience of epileptic seizures, or experience of head surgery). The ethics and safety committees approved the experimental protocol. 

The visual stimuli were presented at the center of a projector screen at 30 Hz. The audio stimuli were presented through MR-compatible noise-cancelling headphones with an appropriate volume level for each participant. For each participant, fMRI data were collected in five separate sessions over five days. Each session consisted of five to six movie-watching runs (each run lasting 564 $\pm$ 75 sec; mean $\pm$ s.t.d.). Five runs (2,823 seconds in total) that contained completely novel music and choreography from the training set were separated as testing runs. As for the clips in the testing dataset, the participants also viewed backward versions of the clips. The participants were given ¥2,500 per hour for scanning to compensate for their time.

We collected fMRI data using a 3.0-Tesla Siemens MAGNETOM Prisma scanner. We performed a T2*-weighted gradient-echo echo planar imaging (EPI) scan to acquire functional images covering the entire brain (TR, 1000 ms; TE, 30 ms; flip angle, 60 deg; voxel size, $2 \times 2 \times 2$ mm; number of slices, 76; multiband factor, 6). We also acquired T1-weighted (T1w) magnetization-prepared rapid acquisition gradient-echo (MP-RAGE) fine-structural images of the entire head (TR, 2500 ms; TE, 2.18 ms; TI, 1000 ms; flip angle, 8 deg; voxel size, $1.0 \times 1.0 \times 1.0$ mm).

We preprocessed the fMRI data using FMRIPREP version 20.1.3 \citep{esteban2019fmriprep}. The preprocessing steps included realignment, slice-time correction, coregistration, segmentation of T1-weighted structural images, and normalization to Montreal Neurological Institute space. For more details of the pipeline, see http://fmriprep.readthedocs.io/en/latest/workflows.html. We also performed an additional nuisance regression step \citep{ciric2017benchmarking} to remove motion parameters and physiological noise. Specifically, we removed six primary motion parameters, along with their derivatives, and the quadratics of all regressors. We modeled physiological noise using white matter, ventricles, and the first five principal components of aCompCor on time series extracted from both white matter and ventricles \citep{behzadi2007component}. Additionally, we included the derivatives of white matter and ventricles, and the quadratics of all physiological noise regressors. The nuisance regression model contained a total of 37 nuisance parameters. We excluded global signal regression (GSR), given that GSR artificially induces negative correlations \citep{murphy2009impact,power2014methods}. We extracted the time series of BOLD signals from the same 10 ROIs and eight large-scale functional networks used in the previous analysis. We shifted the fMRI scans by 6 s to consider the BOLD response.

We used two types of task parameters: motion and audio. For motion, we used 17 COCO-format 3D human joint locations that were provided by the AIST++ Dance Motion Dataset, which resulted in a 51-dimensional motion feature. For audio, we used the publicly available audio processing toolbox Librosa \citep{mcfee2015librosa} to extract the audio features including a 1D envelope, 20-dimensional mel-frequency cepstral coefficient, and 12-dimensional chroma, which resulted in a 33-dimensional music feature. For each dimension of the motion and audio features, we averaged across at 1-s resolution (i.e., BOLD sampling rate) and applied z-score normalization.

We applied RRR, PPI, NDSCA-simple, and NDSCA, and tested the statistical significance as described above.

\bibliography{neurips_2021_spl}
\newpage
\makeatletter
\@fpsep\textheight
\makeatother

\begin{figure}
  \centering
  \includegraphics[width=0.6\linewidth]{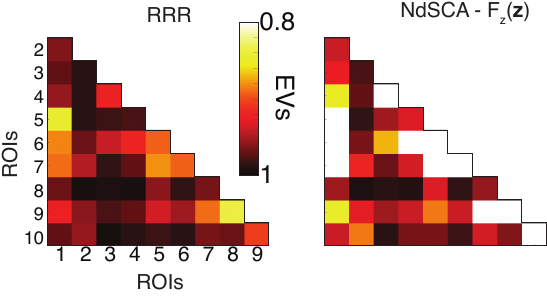}
  \caption{{\bf Cinema movie dataset: demixed interaction associated with latent variable $z$}. The obtained interaction matrices were quite similar overall between non-demixed RRR (left) and our NdSCA with the $F_z(\bf{z})$ term used for demixing (right), except that our method revealed strong connections in some pairs of ROIs more clearly. Detailed analysis and interpretation on this difference are left for our future study.}
\end{figure}
